\documentclass[aps,pra,amsmath,amssymb,twocolumn,groupedaddress,showpacs,eqsecnum]{revtex4-1}
\usepackage{graphicx}
\usepackage[utf8]{inputenc}
\usepackage{amsmath}
\usepackage{epsf}
\newcommand{\RN}[1]{%
  \textup{\uppercase\expandafter{\romannumeral#1}}%
}

\newcommand{\be}{\begin{equation}}
\newcommand{\ee}{\end{equation}}
\newcommand{\ba}{\begin{eqnarray}}
\newcommand{\ea}{\end{eqnarray}}
\newcommand{\baa}{\begin{eqnarray}}
\newcommand{\eaa}{\end{eqnarray}}
\newcommand{\ed}{\end{document}}

\renewcommand{\baselinestretch}{1.2}
\date{\today}

\begin{document}
\title{Fast forward of adiabatic control of tunneling states}
\author{  Katsuhiro Nakamura$^{(1,2)}$,  Anvar Khujakulov$^{(3)}$, Sanat Avazbaev$^{(4)}$ 
and Shumpei Masuda$^{(5)}$}
\affiliation{$^{(1)}$Faculty of Physics, National University of Uzbekistan, Vuzgorodok, Tashkent 100174, Uzbekistan\\
$^{(2)}$Department of Applied Physics, Osaka City University, Sumiyoshi-ku, Osaka 558-8585, Japan\\
$^{(3)}$Institut f\"{u}r Physik, Humboldt-Universit\"{a}t zu Berlin, Newtonstrasse 15, D-12 489 Berlin, Germany\\
$^{(4)}$Faculty of Professional Education, Tashkent State Pedagogical University, 27 Bunyodkor Street, Tashkent 100070, Uzbekistan\\
$^{(5)}$Department of Applied Physics, Aalto University, P.O.Box 13500, FI-00076 AALTO, Finland
}

\begin{abstract}
By developing the preceding work on the fast forward  of transient phenomena of quantum tunneling by Khujakulov and Nakamura (Phys. Rev.  {\bf A 93}, 022101 (2016) ), we propose a scheme of the exact fast forward of adiabatic control of stationary tunneling states with use of  the electromagnetic field. The idea allows the acceleration of both the amplitude and phase of wave functions throughout the fast-forward time range. The scheme realizes the fast-forward observation of the transport coefficients under the adiabatically-changing barrier with the fixed energy of an incoming particle.
As typical examples we choose systems with (1) Eckart's potential with tunable asymmetry and (2) double $\delta$-function barriers under tunable relative height. We elucidate the driving electric field to guarantee the stationary tunneling state during a rapid change of the barrier and evaluate both the electric-field-induced temporary deviation of transport coefficients from their stationary values and the modulation of the phase of complex scattering coefficients.
\end{abstract}
\maketitle

\section{Introduction}\label{eq-introduction}
Various methods to control quantum states have been 
reported in Bose-Einstein condensates (BEC),
quantum computations and many other fields of applied physics.  
It is important to consider the speed-up
of such manipulations of quantum states for manufacturing purposes and
for innovation of technology, because the coherence of systems is degraded by their interaction with the environment.

Masuda and Nakamura \cite{mas1,mas2,mas3} investigated a way to
accelerate quantum dynamics with use of a characteristic driving
potential determined by the additional phase of a wave function.
This kind of acceleration
is called the fast forward, which means to reproduce a series of events or
a history of matters in a shortened time scale, like a rapid
projection of movie films on the screen.

The fast forward theory applied to quantum adiabatic dynamics \cite{mas2,mas3} assumes that a product of
the mean value $\bar{\alpha}$ of an infinitely-large time scaling factor $\alpha(t)$ and an infinitesimally small growth rate $\epsilon$ in the  quasi-adiabatic parameter should satisfy the constraint $\bar{\alpha} \times \epsilon = finite$ in the asymptotic limit $\bar{\alpha}\rightarrow \infty$ and $\epsilon \rightarrow 0$.
The scheme needs no knowledge of spectral properties of the system and is free from the initial and boundary value problem. Therefore it constitutes one of the promising ways of shortcuts to adiabaticity (STA) devoted to tailor excitations in nonadiabatic processes\cite{dr1,dr2,mb,lr,mg1,mg2,campo,djc,mnc}. Some papers \cite{mg-nice,tk-nice} made clear the relationship between 
the fast forward approach and other STA protocols.
Recent interesting application of the fast forward theory can be found in acceleration of Dirac dynamics \cite{deff} and in accelerated construction of classical adiabatic invariant under non-adiabatic circumstances \cite{jarz}.

Although Masuda and Nakamura's  works guarantee the exact target state at the fast-forward final time $t=T_{FF}$, in the intermediate time range $0\le t \le T_{FF}$ they accelerate only the amplitude of the wave function and fail to accelerate its phase because of the non-vanishing additional phase on the way.

Up to now the adiabatic states to be fast forwarded  are limited to bound states. If one wants to accelerate the current-carrying  scattering states, one must innovate the scheme so as to keep the original phase exactly in the intermediate time range until $t=T_{FF}$.

Recently, in the context of the transient phenomena of quantum tunneling,
Khujakulov and Nakamura~\cite{khu} found  a way of fast-forwarding of quantum
dynamics for charged particles by applying the electromagnetic field, which exactly accelerates
both amplitude and phase of the wave function throughout
the fast-forward time range. This means the fast forward with complete fidelity.
The scheme suggests a possibility to accelerate the adiabatic control of stationary scattering  states under the fixed energy of an incoming particle. The scheme of Khujakulov and Nakamura as it stands, however, is not useful and must be innovated so as to be suitable to the adiabatic dynamics characterized by infinitesimally-slowly changing control parameters like the
height and shape of potential barriers.

In this paper we develop the Khujakulov and Nakamura's scheme so that it can be applicable to the fast forward of stationary tunneling states under the adiabatically-changing potential barrier.  To make the paper self-sustained, we shall sketch the general theory of fast forward with complete fidelity~\cite{khu} in Section~\ref{newscheme}. In Section~\ref{FF-adcontrl}, the theory is extended to the fast forward of stationary tunneling dynamics through adiabatically-changing barriers under the fixed energy of an incoming particle. In Section~\ref{trans coefficients} we  show the time-dependent transport coefficients during fast forwarding.
In Section \ref{sample-tunn} typical examples are presented, where we choose systems with  (1) Eckart's potential with tunable asymmetry and
(2) double $\delta$-function barriers with tunable relative height. Conclusion is given in Section~\ref{concl}.
Appendix A is devoted to the gauge transformation of the present scheme to Masuda-Nakamura's one with incomplete fidelity.
Appendix B and C treat the technical details to derive some relevant equations.

 \section{General fast-forward theory with complete fidelity}\label{newscheme}

The Schr\"{o}dinger equation for a charged particle in standard time with a nonlinearity constant $c_0$ (appearing in macroscopic quantum dynamics) is
represented as
\begin{eqnarray}\label{MN-standeq}
\imath\hbar\frac{\partial{\psi_{0}}}{\partial{t}}=-\frac{\hbar^{2}}{2m}\nabla^{2}\psi_{0}
+V_0(\textbf{x},t)\psi_{0} -c_0|\psi_0|^2\psi_0,
\end{eqnarray}
where the coupling with electromagnetic field is assumed to be absent.
$\psi_{0} \equiv \psi_0(\textbf{x},t)$ is a known function of
space $\textbf{x}$ and time $t$ under a given potential $V_0(\textbf{x},t)$ and is called a standard state. For any
long time $T$ called as a standard final time, we choose
$\psi_{0}(t=T)$ as a target state that we are going to generate in a shorter time.

Let $\Lambda(t)$ be the advanced time defined by
\begin{eqnarray}\label{ff-t}
\Lambda(t)=\int_0^t\mathrm{\alpha(t')}\,\mathrm{d}t',
\end{eqnarray}
where $t$ is a time scale shorter than the standard one.
$\alpha (t)$ is a magnification time-scale factor given by
$\alpha(0) =1$, $\alpha(t) >1(0<t<T_{FF})$ and $\alpha(t)=1(t \ge T_{FF})$.
We consider the fast-forward dynamics with a new time variable which reproduces the target state $\psi_0(T)$
in a shorter final time $T_{FF}(<T)$
defined by
\begin{eqnarray}\label{T-Tff}
T=\int_0^{T_{FF}}\alpha (t) \mathrm{d}t.
\end{eqnarray}

The explicit expression for $\alpha(t)$ in the fast-forward range ($0 \le t \le T_{FF}$) is  typically given by \cite{mas1,mas2,mas3} as:
\begin{eqnarray}\label{nonunif-scale}
\alpha(t)=\bar{\alpha}-(\bar{\alpha}-1)\cos\left(\frac{2\pi}{T_{FF}}t\right),
\end{eqnarray}
where $\bar{\alpha}$ is the mean value of $\alpha(t)$ and is given by $\bar{\alpha}=T/T_{FF}$.
Besides the time-dependent scaling factor in Eq.(\ref{nonunif-scale}) in the fast-forward time range, we can also have recourse to the uniform scaling factor $\alpha(t)=\bar{\alpha}  ( 0\le t \le T_{FF})$,
which is useful in the quantitative analysis of fast forward.

The fast-forward wave function $\psi_{FF}$ in this paper does not include the additional phase
and is given by
\begin{eqnarray}\label{ffwf}
\psi_{FF}(\textbf{x},t)=\psi_{0}(\textbf{x},\Lambda(t))\equiv\tilde{\psi}_{0}(\textbf{x},t).
\end{eqnarray}
$\psi_{FF}$ is just like a movie film projected on the screen in a shortened time scale. 
Equation (\ref{ffwf}) guarantees the complete fidelity, namely $\langle  \psi_{FF}|\tilde{\psi_0} \rangle =1$ throughout the fast forward time range. We shall realize $\psi_{FF}$ by applying the electromagnetic field $\textbf{E}_{FF}$
and $\textbf{B}_{FF}$ which are related  to vector $\textbf{A}_{FF}(\textbf{x},t)$ and scalar
$V_{FF}(\textbf{x},t)$ potentials through
\begin{align}\label{EB-AV}
\begin{split}
\textbf{E}_{FF}&=-\frac{\partial \textbf{A}_{FF}}{\partial t}-\nabla V_{FF}, \\
\textbf{B}_{FF}&=\nabla\times\textbf{A}_{FF}.
\end{split}
\end{align}

Let's assume $\psi_{FF}$ to be the solution of the Schr\"{o}dinger
equation for a charged particle in the presence of
$\textbf{A}_{FF}(\textbf{x},t)$ and
$V_{FF}(\textbf{x},t)$, as given by
%
\begin{eqnarray}\label{ffeq}
\imath\hbar\frac{\partial{\psi_{FF}}}{\partial{t}}&=&\hat{H}_{FF}\psi_{FF}\nonumber\\
&\equiv&\left(\frac{1}{2m}\left(\frac{\hbar}{i}\nabla-\frac{q}{c}\textbf{A}_{FF}\right)^2+ q V_{FF}
+V_0\right)\psi_{FF} \nonumber\\
&-& c_0|\psi_{FF}|^2\psi_{FF}.
\end{eqnarray}
For simplicity, we shall hereafter employ the unit  velocity of light $c=1$ and the prescription of a positive unit charge $q=1$.  $V_{FF}$
in Eq.(\ref{ffeq}) is introduced independently from a given potential $V_0$, in contrast to the preceding work \cite{mas1}.  
The electromagnetic field investigated in Refs. \cite{mas3,kiel} was not used to suppress the additional phase.

Replacing $t$ by $\Lambda(t)$ in Eq.(\ref{MN-standeq}) and noting Eq.(\ref{ffwf}),
we can eliminate $\frac{\partial \tilde{\psi}_{0}}{\partial t}$ between Eqs.(\ref{MN-standeq}) and (\ref{ffeq}).
The resultant equality is decomposed into real and imaginary parts as respectively given by 
\begin{eqnarray}\label{aff}
\nabla\cdot\textbf{A}_{FF}&+&2\textrm{Re}\left[\frac{\nabla\tilde{\psi}_{0}}{\tilde{\psi_{0}}}\right]\cdot\textbf{A}_{FF} \nonumber\\
&+&\hbar(\alpha-1)
\textrm{Im}\left[\frac{\nabla^{2}\tilde{\psi}_{0}}{\tilde{\psi_{0}}}\right]=0
\end{eqnarray}
and
\begin{align}\label{vff}
\begin{split}
V_{FF}&=-(\alpha-1)\frac{\hbar^{2}}{2m}\textrm{Re}\left[\frac{\nabla^{2}\tilde{\psi}_{0}}{\tilde{\psi_{0}}}\right] 
+\frac{\hbar}{m}\textbf{A}_{FF}\cdot\textrm{Im}\left[\frac{\nabla\tilde{\psi}_{0}}{\tilde{\psi_{0}}}\right] \\
&-\frac{1}{2m}\textbf{A}_{FF}^{2}+(\alpha-1)V_0 
-(\alpha-1)c_0|\tilde{\psi_0}|^2.
\end{split}
\end{align}

Rewriting
$\tilde{\psi}_{0}$ in terms of the real positive amplitude $\rho$ and phase $\eta$ as
\begin{eqnarray}
\tilde{\psi}_{0}=\rho(\textbf{x},\Lambda(t)) {\rm exp}(i\eta(\textbf{x},\Lambda(t))),
\end{eqnarray}
we find that
\begin{eqnarray}\label{aff-sol}
\textbf{A}_{FF}=-\hbar(\alpha-1)\nabla\eta
\end{eqnarray}
satisfies Eq.(\ref{aff}).
Using Eq.(\ref{aff-sol}), $V_{FF}$ can be expressed only in terms of $\eta$ as
\begin{align}\label{vff-sol}
\begin{split}
V_{FF}&=-(\alpha-1)\hbar\frac{\partial \eta}{\partial \Lambda(t)}
-\frac{\hbar^{2}}{2m}(\alpha^2-1)(\nabla \eta)^2.
\end{split}
\end{align}
Applying the driving vector
$\textbf{A}_{FF}$ and scalar $V_{FF}$ potentials in
Eqs.(\ref{aff-sol}) and (\ref{vff-sol}), we can realize the fast-forwarded state $\psi_{FF}$ in Eq.(\ref{ffwf})
which is now free from the additional phase $f$  used in Ref.\cite{mas1}.

Two points should be noted:
1) The above driving potentials do not explicitly depend on the nonlinearity coefficient $c_0$:
Eqs.(\ref{aff-sol}) and (\ref{vff-sol}) work for the nonlinear Schr\"odinger equation  as well;
2) The magnetic field ${\bf B}_{FF}$ is vanishing, because a combination of Eqs. (\ref{EB-AV})
and (\ref{aff-sol}) leads to
${\bf B}_{FF}={\bf \nabla} \times {\bf A}_{FF}=0$. Therefore, only the electric field
$\textbf{E}_{FF}$ is required to accelerate  a given dynamics. With use of
Eqs. (\ref{EB-AV}), (\ref{aff-sol}) and (\ref{vff-sol}), we find:
$\textbf{E}_{FF}=\hbar \dot{\alpha}\nabla \eta+ \hbar \frac{\alpha^2-1}{\alpha}\partial_t\nabla\eta +\frac{\hbar^{2}}{2m}(\alpha^2-1)\nabla (\nabla \eta)^2$.

A remarkable issue of the present scheme is the enhancement of the current density $\textbf{j}_{FF}$. Using a generalized momentum which includes a contribution from the vector potential in
Eq.(\ref{aff-sol}), we see:
\begin{eqnarray}\label{ff-curren}
\textbf{j}_{FF}(\textbf{x},t)&\equiv&
\textrm{Re}[\psi_{FF}^{*}(\textbf{x},t)\frac{1}{m}\left(\frac{\hbar}{i}\nabla - \textbf{A}_{FF}\right)
\psi_{FF}(\textbf{x},t)]\nonumber\\
&=&\frac{\hbar}{m}\alpha(t)\rho^2(\textbf{x},\Lambda(t))\nabla\eta(\textbf{x},\Lambda(t))\nonumber\\
&=&\alpha(t)\textbf{j}(\textbf{x}, \Lambda(t))
\end{eqnarray}
under the prescription of a positive unit charge, where
the current density in the standard dynamics is defined by
$\textbf{j}(\textbf{x},t) \equiv
\textrm{Re}[\psi_0^{*}(\textbf{x},t)\frac{\hbar}{i m}\nabla
\psi_0(\textbf{x},t)]
= \frac{\hbar}{m}\rho^2(\textbf{x},t)\nabla\eta(\textbf{x},t)$.
Thus the standard current density becomes both time-squeezed and magnified by a time-scaling factor
$\alpha(t)$  in Eq.(\ref{nonunif-scale})  as a result of the exact fast forwarding of wave function throughout the time evolution.
The present scheme is applicable to the fast forward of diverse quantum-mechanical phenomena.

\renewcommand{\labelenumi}{\theenumi.}

\section{Fast forward of adiabatic change of  tunneling states}\label{FF-adcontrl}
Section \ref{newscheme} was concerned with the fast forward of standard dynamics with standard time scale. From now on, we shall investigate the fast forward of very slow dynamics, i.e., of quasi-adiabatic dynamics. Confining to 1 dimensional ($1D$) system and suppressing the nonlinear term proportional to $c_0$, we shall apply the scheme in Section \ref{newscheme} to stationary tunneling states under an adiabatically-changeable potential barrier, and show the fast forward of adiabatic control of $1D$ tunneling
states with use of the electromagnetic field. The goal of this Section is to obtain the driving gauge potentials and electric field to guarantee such fast forwarding.

We shall take the following strategy: (i) A given potential barrier $V_{0}$ is assumed to 
change adiabatically, and we find a stationary state $\psi_{0}$, which is a solution
of the time-independent Schr\"{o}dinger equation with the instantaneous Hamiltonian; (ii)
Then both $\psi_{0}$ and $V_{0}$ are regularized so that they should satisfy the time-dependent Schr\"{o}dinger equation; (iii) Finally, taking the regularized state as a standard state, we apply the scheme in Section \ref{newscheme}, where the mean value $\bar{\alpha}$ of the infinitely-large time scaling factor $\alpha(t)$ will be chosen to cope with the infinitesimally-small growth rate $\epsilon$ of the quasi-adiabatic parameter and to satisfy $\bar{\alpha} \times \epsilon=finite$.

Let's consider the standard dynamics with a potential barrier characterized by a slowly-varying control parameter $R(t)$ given by
\begin{eqnarray}\label{eq3.1}
R(t)=R_{0}+\epsilon t,
\end{eqnarray}
with the growth rate $\epsilon\ll1$, which means that it requires a very long time $T=O\left(\frac{1}{\epsilon}\right)$, to see the recognizable change of $R(t)$. The time-dependent $1D$ Schr\"{o}dinger equation without the nonlinear term  is:
\begin{eqnarray}\label{eq3.2}
\textit{i}\hbar\frac{\partial\psi_{0}}{\partial t}=-\frac{\hbar^{2}}{2m}\partial_x^{2}\psi_{0}+V_{0}(x,R(t))\psi_{0}.
\end{eqnarray}
The stationary tunneling state $\phi_{0}$ satisfies 
the time-independent counterpart given by
\begin{eqnarray}\label{eq3.3}
E\phi_{0}=\hat{H}_0\phi_{0}\equiv\left[-\frac{\hbar^{2}}{2m}\partial_x^{2}+V_{0}(x,R)\right]\phi_{0}.
\end{eqnarray}

Without loss of generality, we assume that $V_{0}(x,R)$ is $R$-independent constant for $x\le x_1$ and $x\ge x_2$ and shows a $R$-dependent variation for $x_1 \le x \le x_2$. 
In fact, potential barriers are adiabatically controllable in a finite spatial region.

In case of the bound states, the boundary condition for $\phi_{0}$ is $\phi_{0}\rightarrow 0$ at $|x| \rightarrow \infty$, giving the discrete energy spectra. In case of scattering states which includes tunneling states, however, an arbitrary one of the continuum energy is first given, which then determines the stationary scattering state. 

Here we investigate the following situation:
(1) The potential barrier $V_{0}(x,R)$ is deformed very slowly through 
the adiabatic parameter $R$;
(2) During the above adiabatic deformation of $V_{0}(x,R)$, the energy of a plane-wave type particle incoming from the left is assumed to be
$R$-independent and fixed, i.e.,
\begin{equation}\label{fixE}
\frac{\partial E}{\partial R}=0.
\end{equation}

Then, with use of the stationary tunneling state $\phi_0$ satisfying Eq.(\ref{eq3.3}),
one might conceive the corresponding time-dependent state to be a product of $\phi_0$ and a dynamical factor as,
\begin{eqnarray}\label{eq3.4}
\psi_{0}=\phi_{0}(x,R(t))e^{-\frac{\textit{i}}{\hbar}Et}.
\end{eqnarray}
However, $\psi_{0}$ as it stands does not satisfy Eq.(\ref{eq3.2}). Therefore we introduce a regularized state
\begin{eqnarray}\label{eq3.5}
\psi^{reg}_{0}
&\equiv& \phi_{0}(x,R(t))e^{\textit{i}\epsilon\theta(x,R(t))}e^{-\frac{i}{\hbar}Et}
\nonumber\\
&\equiv& \phi_0^{reg} (x,R(t))e^{-\frac{i}{\hbar}Et}
\end{eqnarray}
together with a regularized potential
\begin{eqnarray}\label{eq3.6}
V^{reg}_{0}\equiv V_{0}(x,R(t))+\epsilon\tilde{V}(x,R(t)).
\end{eqnarray}
$\theta$ and $\tilde{V}$ will be determined self-consistently so that $\psi^{reg}_{0}$ should fulfill the time-dependent Schr\"{o}dinger equation,
\begin{eqnarray}\label{eq3.7}
\textit{i}\hbar\frac{\partial\psi_{0}^{reg}}{\partial t}=-\frac{\hbar^{2}}{2m}\partial_x^{2}\psi^{reg}_{0}+V^{reg}_{0}\psi^{reg}_{0},
\end{eqnarray}
up to the order of $\epsilon$.

Rewriting $\phi_{0}(x,R(t))$ with use of the real positive amplitude $\overline{\phi}_{0}(x,R(t))$ and phase $\eta(x,R(t))$ as
\begin{eqnarray}\label{eq3.8}
\phi_{0}(x,R(t))=\bar{\phi}_{0}(x,R(t))e^{\textit{i}\eta(x,R(t))},
\end{eqnarray}
we see $\theta$ and $\widetilde{V}$ to satisfy:
\begin{eqnarray}\label{eq3.9}
\partial_x(\bar{\phi}_{0}^2\partial_x\theta) =-\frac{m}{\hbar}\partial_{R}\bar{\phi}_{0}^2,
\end{eqnarray}
\begin{eqnarray}\label{eq3.10}
\frac{\tilde{V}}{\hbar}=-\partial_{R}\eta-\frac{\hbar}{m}\partial_x\eta\cdot\partial_x\theta.
\end{eqnarray}
Integrating Eq. (\ref{eq3.9}) over  $x$, we have
\begin{equation}\label{dtheta}
\partial_x\theta=-\frac{m}{\hbar}\frac{1}{\bar{\phi}_0^2}\int_{c}^x \partial_R\bar{\phi}_0^2 dx', 
\end{equation}
with $c$ an arbitrary $R$-independent constant. Equation (\ref{dtheta}) determines $\tilde{V}$ in Eq.(\ref{eq3.10}).

In the stationary (or steady) scattering state, the current density available from Eqs.(\ref{eq3.4}) with (\ref{eq3.8}), 
\begin{equation}\label{stcurr}
\textrm{Re}\left[\psi_0^*\frac{\hbar}{im}\partial_x\psi_0\right]=
\frac{\hbar}{m}\bar{\phi}^{2}_{0}(x,R)\partial_x\eta(x,R),
\end{equation}
is space-independent and non-zero constant. Therefore, $\bar{\phi}_{0}$ cannot be zero and the right-hand side of Eq.(\ref{dtheta})
is free from the problem of wave function nodes proper to excited states of bound systems. See also Appendix A.

Applying the scheme in Section \ref{newscheme}, we shall take $\psi^{reg}_{0}$ as a standard state and define its fast-forward version 
$\psi_{FF}$ as
\begin{eqnarray}\label{eq3.11}
\psi_{FF}(x,t)
&\equiv& \phi^{reg}_{0}(x,R(\Lambda(t)))e^{-\frac{i}{\hbar}Et}\nonumber\\
&\equiv& \tilde{\phi}^{reg}_{0}(x,t)e^{-\frac{i}{\hbar}Et}.
\end{eqnarray}
$\psi_{FF}(x,t)$ is then assumed to obey the time-dependent Schr\"{o}dinger equation for a charged particle in the presence of electromagnetic field, as in Eq.(\ref{ffeq}). Then $ \tilde{\phi}^{reg}_{0}(x,t)$ satisfies
\begin{eqnarray}\label{eq3.12}
\textit{i}\hbar\frac{\partial\tilde{\phi}^{reg}_{0}}{\partial t}&=&\frac{1}{2m}\left(\frac{\hbar}{\textit{i}}\partial_x-\frac{q}{c}A_{FF}\right)^{2}\tilde{\phi}^{reg}_{0} \nonumber\\
&+&(qV_{FF}+V_{0}-E+\epsilon\tilde{V})\tilde{\phi}^{reg}_{0},
\end{eqnarray}
where $A_{FF}$ and $V_{FF}$ are gauge potentials to guarantee the exact fast forward.
Here $V_{0}\equiv V_{0}(x,R(\Lambda(t)))$
and $\tilde{V}\equiv\tilde{V}(x,R(\Lambda(t)))$. The dynamical phase in Eq.(\ref{eq3.11}) has led to
the energy shift in the potential in Eq.(\ref{eq3.12}).

In the context of the fast forward of the adiabatic control, it is essential to analyze equalities in Eqs.(\ref{aff}) and (\ref{vff}) directly, because $\tilde{\psi}_{0}$ and $V_0$ there should now be read as
\begin{eqnarray}\label{eq3.13}
\tilde{\psi}_{0}&\rightarrow&\tilde{\phi}^{reg}_{0}\nonumber\\
&\equiv & \bar{\phi}_{0}(x,R(\Lambda(t)))e^{\textit{i}\left[\eta(x,R(\Lambda(t)))
+\epsilon\theta(x,R(\Lambda(t)))\right]} \nonumber\\
\end{eqnarray}
and
\begin{eqnarray}\label{V-tilde}
V_{0} \rightarrow V_0-E+\epsilon\tilde{V},
\end{eqnarray}
respectively. Then Eqs. (\ref{aff}) and (\ref{vff}) lead to the driving $A_{FF}$ and $V_{FF}$ potentials to realize the fast-forward state
$\psi_{FF}$ in Eq.(\ref{eq3.11}):
\begin{eqnarray}\label{eq3.15}
A_{FF}=-\hbar\epsilon(\alpha-1)\partial_x\theta 
\end{eqnarray}
and
\begin{eqnarray}\label{eq3.16}
V_{FF}&=&-\frac{\hbar^{2}}{m}\epsilon(\alpha-1)\partial_x\theta\cdot\partial_x\eta \nonumber\\
&-&\alpha(\alpha-1)\frac{\hbar^{2}}{2m}\epsilon^{2}(\partial_x\theta)^{2}-\epsilon(\alpha-1)\hbar\partial_{R}\eta. \nonumber\\
\end{eqnarray}
The derivation of Eqs. (\ref{eq3.15}) and (\ref{eq3.16}) is given in Appendix B.

Now, applying our central strategy to take the limit $\epsilon\rightarrow 0$ and $\bar{\alpha}\rightarrow \infty$ with $\epsilon\bar{\alpha}=\bar{v}$ being kept finite, we can reach the issue:
\begin{eqnarray}\label{eq3.17}
A_{FF}&=&-\hbar v(t)\partial_x\theta,\nonumber\\
V_{FF}&=&-\frac{\hbar^{2}}{m}v(t)\partial_x\theta\cdot\partial_x\eta \nonumber\\
&-&\frac{\hbar^{2}}{2m}(v(t))^{2}(\partial_x\theta)^{2}-\hbar v(t)\partial_{R}\eta,
\end{eqnarray}
where, with use of $T_{FF}\left(=\frac{T}{\bar{\alpha}}=O\left(\frac{1}{\epsilon\bar{\alpha}}\right)\right)=finite$,
%
\begin{eqnarray}\label{accel-ad}
v(t)&\equiv&\lim_{\epsilon\rightarrow 0, \bar{\alpha}\rightarrow \infty}\varepsilon\alpha(t)=\bar{v}\left(1-\cos\frac{2\pi}{T_{FF}}t\right),\nonumber\\
R(\Lambda(t))&=&R_0+\lim_{\epsilon\rightarrow 0, \bar{\alpha}\rightarrow \infty}\varepsilon\Lambda(t)\nonumber\\
&=& R_{0}+\bar{v}\left(t-\frac{T_{FF}}{2\pi}\sin\left(\frac{2\pi}{T_{FF}}t\right)\right),\nonumber\\
&& \;{\rm for} \; 0  \le t \le T_{FF},
\end{eqnarray}
and
\begin{eqnarray}\label{beyon}
v(t)=0,  \quad R(\Lambda(t))=R_0+\bar{v}T_{FF}  \quad {\rm for} \; T_{FF} \le t.\nonumber\\
\end{eqnarray}
$v(t)$ and its mean $\bar{v}$ stand for the time-scaling factors coming from $\alpha(t)$ and $\bar{\alpha}$, respectively.

In the same limiting case as above, $\psi_{FF}$ is explicitly given by
\begin{eqnarray}\label{eq3.19}
\psi_{FF}=\bar{\phi}_{0}(x,R(\Lambda(t)))
e^{\textit{i}\eta(x,R(\Lambda(t)))}e^{-\frac{i}{\hbar}Et},
\end{eqnarray}
and describes the acceleration of the adiabatic control of stationary scattering states  throughout the fast forward time range until $t\leq T_{FF}$. It should be emphasized: while $\bar{\alpha} \rightarrow +\infty$ is assumed, the gauge potential and electromagnetic field are 
of  finite order (i.e., $O(\bar{v})$ or $O(\bar{v}^2)$).

>From Eq.(\ref{eq3.17}),  the driving electric field to guarantee the fast-forward state in Eq.(\ref{eq3.19}) is given by
\begin{eqnarray}\label{eq3.24}
E_{FF}&=&-\frac{\partial A_{FF}}{\partial t}-\partial_x V_{FF}\nonumber\\
 &=&\hbar\dot{v}\partial_x\theta + \hbar v^2(t)\partial_R(\partial_x\theta) +\frac{\hbar^{2}}{2m}v(t)\partial_x(\partial_x\theta\cdot\partial_x\eta)\nonumber\\
 &+&\frac{\hbar^{2}}{2m}(v(t))^{2}\partial_x(\partial_x\theta)^{2}+\hbar v\partial_{R}(\partial_x\eta).\end{eqnarray}
In SI unit for electric field, our dimensionless $E_{FF}$ corresponds to
$E^{FF}_{SI}=\frac{m_e c \omega}{e}\times E_{FF}\sim\frac{10^6}{\lambda} E_{FF}$ where
$m_e, e, c, \omega$ and $\lambda$ are electron mass, electron charge, velocity of light, frequency of laser light and its wave length, respectively. Typical value $E_{FF}=1$ in case of IR lasers of wave length $\sim$ 1$\mu$m means $E^{FF}_{SI}=10^{12}$. 

Note: (1) We need the space-(and time-)dependent electric field $E_{FF}$ along the $1D$ target system on $x$-axis, which means that $\partial_x E_{FF}$ is nonzero. On the other hand, the Maxwell's equation (Gauss's law) requires the divergence of electric field $=\partial_x E_x + \partial_y E_y + \partial_z E_z=$ charge density.  The experimental setup to be compatible with the Maxwell's equation is to apply the electric field (surrounding the target system) which has 3 components and exists in $3D$ space, so that  the perpendicular components $(E_y, E_z)$ should satisfy $ \partial_y E_y + \partial_z E_z =- \partial_x E_x(\equiv - \partial_x E_{FF})$ along the $x$-axis. An example is to prepare an infinite straight rod which is detached from and perpendicular to the target system and to introduce the inhomogeneous charge distribution along the rod so that 
$E_x = E_{FF}$ should appear along the $x$-axis. In this case, no charge distribution is necessary along the target system. (2) The time-dependent electric field might induce a magnetic $B$ field due to the
Ampere-Maxwell's equation. Since we are concerned with $1D$ tunneling and the electric field is applied along the $x$ direction, such $B$ field  
is perpendicular to $x$-axis, and the Lorentz force working on the target particle is perpendicular to both $x$-axis and the direction of $B$ field. Therefore, $B$ field plays no role in the tunneling along $x$-axis.

In closing this Section, we should note: the scheme here is the theory of  fast forward with complete fidelity, but is compatible with that of the preceding one with the additional phase \cite{mas2,mas3}, as proved by using the gauge transformation in Appendix \ref{app1}.

\section{Fast forward of observation of adiabatically-tunable transport coefficients}\label{trans coefficients}
Now we shall elucidate the time-dependent transport (i.e., transmission and reflection) coefficients during the accelerated adiabatic control of stationary tunneling states.

With use of the results in Eqs. (\ref{eq3.17}) and (\ref{eq3.19}), the current density $j_{FF}$ during the fast forward time region becomes
\begin{eqnarray}\label{eq3.25}
j_{FF}(x,t)&=&\textrm{Re}\left[\psi_{FF}^{*}(x,t)\frac{1}{m}
 \left(\frac{\hbar}{\textit{i}}\partial_x - A_{FF}\right)\psi_{FF}(x,t)\right]\nonumber\\
 &=& j_{ad}(x,t)+j_{nad}(x,t),
\end{eqnarray} 
where
\begin{equation}\label{curr-ad}
j_{ad}(x,t)\equiv \frac{\hbar}{m}\bar{\phi}^{2}_{0}(x,R(\Lambda(t)))\partial_x\eta(x,R(\Lambda(t)))
\end{equation}
and
\begin{eqnarray}\label{curr-nonad}
j_{nad}(x,t)&\equiv &  v(t)\frac{\hbar}{m}\bar{\phi}^{2}_{0}(x,R(\Lambda(t)))\partial_x\theta(x,R(\Lambda(t)))
\nonumber\\
&=& -v(t)\int_{c}^x \partial_R\bar{\phi}_0^2 dx'.
\end{eqnarray}
The last equality of Eq.(\ref{curr-nonad}) comes from Eq.(\ref{dtheta}). The decomposition of $ j_{FF}$ into two parts as in Eq.(\ref{eq3.25}) 
was not seen in the fast forward of the standard dynamics in Section 2.
The adiabatic current $j_{ad}$ guarantees transmission and reflection coefficients to precisely reproduce the stationary values during the period of fast forward because of the complete fidelity of $\psi_{FF}(x,t)$.
On the other hand, the nonadiabatic current $j_{nad}$ caused by the driving electric field $E_{FF}(t)$ in Eq. (\ref{eq3.24}) vanishes at the end of fast forward.

The adiabatic potential barrier $V_0(x, R(t))$ is characterized by a slowly-varying control parameter $R(t)$ in Eq.(\ref{eq3.1}). 
As noted in the previous Section, we shall choose $V_0=0$ and $V_0=V_0^c$ ($R$-independent constant) for $x \le x_1$ and $x \ge x_2$, respectively, assuming that the $R$-dependent barrier exists 
only in the range $x_1 \le x \le x_2$.

Before reaching the formula for time-dependent transport coefficients, we shall sketch the stationary state and show the time-independent transport coefficients in $1 D$ systems with the barrier in the adiabatic limit $R(t)=R$=constant.
For the electron with $R$-independent energy $E$  incoming from the left, the  wave function for $x\le x_1$ and 
$x\ge x_2$ is given respectively by
\begin{eqnarray}\label{leftWF}
\psi_{0}&=&\left(e^{ikx}+r_f(R)e^{-ikx}\right)e^{-\frac{i}{\hbar}Et},
\end{eqnarray}
and
\begin{eqnarray}\label{rightWF}
\psi_{0}=t_r(R)e^{ik^{\prime}x}e^{-\frac{i}{\hbar}Et}.
\end{eqnarray}
Here both $k=\frac{1}{\hbar}\sqrt{2mE}$ and $k^{\prime}=\frac{1}{\hbar}\sqrt{2m(E-V_0^c)}$ are $R$-independent constants. $t_r(R)$ and $r_f(R)$ mean the $R$-dependent transmission and reflection coefficients, respectively.

The current densities at $x= x_2$ and $x = x_1$ are:
\begin{eqnarray}\label{lim-curr}
j(x= x_2, R)&=&\textrm{Re}\left[\psi_0^*\frac{\hbar}{im}\partial_x\psi_0\right]\nonumber\\
&=&
\frac{\hbar k'}{m}|t_r(R)|^2 \equiv j_t(R),\nonumber\\
j(x= x_1, R)&=&\frac{\hbar k}{m}(1-|r_f(R)|^2)) \nonumber\\
&\equiv& j_0-j_r(R),
\end{eqnarray}
where
\begin{eqnarray}\label{in-curr}
j_0 \equiv \frac{\hbar k}{m}
\end{eqnarray}
is $R$-independent  fixed current of the incoming particle.
The transmission and reflection probabilities are given by
\begin{eqnarray}\label{transPR}
\mathcal{T}(k,R)=\frac{j_t(R)}{j_0}=\frac{k'}{k}|t_r(R)|^2
\end{eqnarray}
and
\begin{eqnarray}\label{reflePR}
\mathcal{R}(k,R)=\frac{j_r(R)}{j_0}=|r_f(R)|^2,
\end{eqnarray}
respectively. In the stationary state, the current density is space-independent and one can assume
$j(x= x_2, R)=j(x= x_1, R)$. Then we see $j_t(R) + j_r(R)=j_0$ and thereby the unitarity condition
\begin{equation}\label{inst-unitari}
\mathcal{T}(k,R)+\mathcal{R}(k,R)=1 
\end{equation}
for any value of $R$.

Now, consider the fast forward of adiabatic change of the potential barrier 
under the injection of $R$-independent fixed current density $j_0$. 
Then Eqs.(\ref{eq3.25}), (\ref{curr-ad})
and (\ref{curr-nonad})  lead to the  current densities on $x= x_2$ and $x = x_1$ at arbitrary time $t$:
\begin{align}\label{lim-FFcurr}
\begin{split}
j_{FF}(x=x_2, t)&=j_t(R)-v(t)\int_{c}^{x_2} \partial_R\bar{\phi}_0^2 dx, \\
j_{FF}(x=x_1, t)&=j_0-j_r(R)-v(t)\int_{c}^{x_1} \partial_R\bar{\phi}_0^2 dx,
\end{split}
\end{align}
where the accelerated adiabatic parameter
$R \equiv R(\Lambda(t))$ and the time scaling factor $v(t)$ are given in Eqs.(\ref{accel-ad}) and (\ref{beyon}), respectively. By dividing the relevant part of Eq. (\ref{lim-FFcurr}) by $j_0$, we obtain the formula for the time-dependent transmission and reflection coefficients:\begin{eqnarray}\label{FFtransPR}
\mathcal{T}_{FF}(k,t)&\equiv&\frac{j_{FF}(x=x_2, t)}{j_0} \nonumber\\
&=&\mathcal{T}(k,R)-\frac{m}{\hbar k}v(t)\int_{c}^{x_2} \partial_R\bar{\phi}_0^2 dx,
\end{eqnarray}
and
\begin{eqnarray}\label{FFreflPR}
\mathcal{R}_{FF}(k,t)&\equiv&\frac{j_0 - j_{FF}(x=x_1, t)}{j_0} \nonumber\\
&=&\mathcal{R}(k,R)+\frac{m}{\hbar k}v(t)\int_{c}^{x_1} \partial_R\bar{\phi}_0^2 dx,
\end{eqnarray}
respectively.  Equations (\ref{FFtransPR}) and (\ref{FFreflPR}) are the goal of this Section.

The fast forward of adiabatic change of  the stationary tunneling state is actually non-stationary dynamics, and Eqs. (\ref{FFtransPR}) and (\ref{FFreflPR})  together with Eq. (\ref{inst-unitari}) lead to the condition:
\begin{eqnarray}\label{FF-nonunitari}
\mathcal{T}_{FF}(k,t)+\mathcal{R}_{FF}(k,t)
&=&1-\frac{m}{\hbar k}v(t)\int_{x_1}^{x_2} \partial_R\bar{\phi}_0^2 dx \nonumber\\
&\equiv& 1+\delta u.
\end{eqnarray}
The nonadiabatic correction on the right-hand side of Eq.(\ref{FF-nonunitari}), which is $c$-independent, 
shows a deviation $\delta u$ from the unitarity and vanishes at $t=T_{FF}$.
The analysis of the continuity equation of the fast-forward dynamics can also reproduce Eq.(\ref{FF-nonunitari}) (see Appendix C).

The transport coefficients described above are actually transport probabilities.
The stationary states at $x \le x_1$ and $x \ge x_2$ can also be characterized by complex scattering coefficients $r_f(R)$ and  $t_r(R)$ as in Eqs.(\ref{leftWF}) and (\ref{rightWF}).
If one wishes to know the deviation of their phase during the fast forward time, it is convenient to construct the $A_{FF}$-field(gauge-field)-free variant of the present theory of fast forward. This can be done by using the Gauge transformation as in Appendix A. Then $\psi_{FF}$ in Eq.(\ref{eq3.19}) acquires the phase which compensates the $A_{FF}$ field, and becomes:
\begin{eqnarray}\label{free-psi}
\psi_{FF}^{MN}&=&\bar{\phi}_{0}(x,R(\Lambda(t)))
e^{\textit{i}\eta(x,R(\Lambda(t)))}\nonumber\\
&\times&e^{\textit{i}v(t)\theta(\textbf{x},R(\Lambda(t)))}e^{-\frac{i}{\hbar}Et}.
\end{eqnarray}
At  $x \ge x_2$ where $V_0(x,R)$ is $R$-independent constant, noting $\bar{\phi}_{0}e^{\textit{i}\eta}=t_r(R)e^{ik'x}$, the fast-forward variant of Eq.(\ref{rightWF}) becomes:
\begin{eqnarray}\label{free-trans}
\psi_{FF}^{MN}=t_r^{FF}(R(\Lambda(t)))
e^{ik'x} e^{-\frac{i}{\hbar}Et} 
\end{eqnarray}
with
\begin{eqnarray}\label{free-trans-coff}
t_r^{FF}(R(\Lambda(t)))=t_r(R(\Lambda(t)))
e^{\textit{i}v(t)\theta(x_2,R(\Lambda(t)))}.
\end{eqnarray}
The $A_{FF}$-field-free current density at $x=x_2$  is  calculated in the same way as in Eq.(\ref{lim-curr}) and is given by
\begin{eqnarray}\label{free-trans-curr}
j_{FF}^{MN}(x= x_2, t)&=&\textrm{Re}\left[\psi_{FF}^{MN*}\frac{\hbar}{im}\partial_x\psi_{FF}^{MN}\right]_{x=x_2}\nonumber\\
&=&
\frac{\hbar k'}{m}|t_r(R(\Lambda(t)))|^2 +v(t)\frac{\hbar}{m}\bar{\phi}_0^2 \partial_x \theta |_{x=x_2}.\nonumber\\
\end{eqnarray}
Recalling the formula in Eq.(\ref{dtheta}),
Eq.(\ref{free-trans-curr}) proves to be equal to Eq.(\ref{lim-FFcurr}), and, after its scaling by $j_0$ in Eq.(\ref{in-curr}), exactly reproduces the time dependent transport coefficients in Eq.(\ref{FFtransPR}).
The shoulder of the exponential of $t_r^{FF}$ in Eq.(\ref{free-trans-coff}) represents the phase modulation of scattering coefficients
during the fast forward time, and, with use of Eq.(\ref{dtheta}), is explicitly given by
\begin{eqnarray}\label{phase-dev}
v(t)\theta(x_2,R(\Lambda(t)))=-v(t)\frac{m}{\hbar} \int_{c}^{x_2} \frac{dx}{\bar{\phi}_0^2} \int_{c}^{x} \partial_R\bar{\phi}_0^2 dx'. \nonumber\\
\end{eqnarray}
Since $\bar{\phi}_0$ has no nodes as explained below Eq. (\ref{stcurr}),  the double integrals in Eq.(\ref{phase-dev}) is finite and  
the phase $v(t)\theta(x_2,R(\Lambda(t)))$ vanishes at the end of the fast forward.
Similarly, the fast-forward variant of $r_f(R)$ is given by
\begin{eqnarray}\label{free-refl-coff}
r_f^{FF}(R(\Lambda(t)))=r_f(R(\Lambda(t)))
e^{\textit{i}v(t)\theta(x_1,R(\Lambda(t)))},
\end{eqnarray}
where the expression for $v(t)\theta(x_1,R(\Lambda(t)))$ is given by Eq.(\ref{phase-dev}) 
with the upper integration limit $x_2$ replaced by $x_1$.

The important finding in this Section is that, throughout the fast forward time range the transport coefficients include the nonadiabatic contribution, which vanishes at the goal when
$v(t)=0$, namely both  $\mathcal{T}_{FF}(k,t)$ and $\mathcal{R}_{FF}(k,t)$ exactly reproduce the adiabatic counterparts
at the end of the fast forward.

\begin{figure}[htb]\label{V-pf}
\centering
\includegraphics[width=1.0\linewidth]{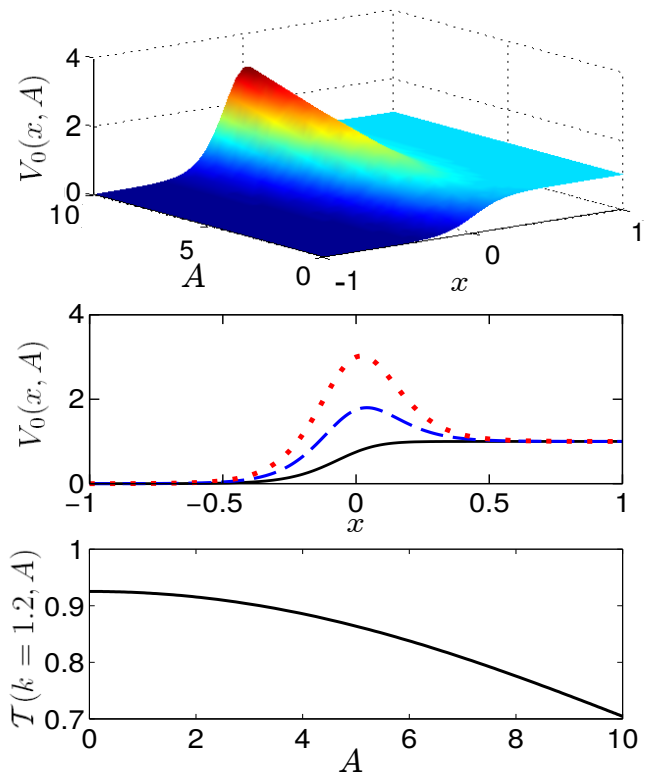}
\caption{Upper two panels: Eckart's potential in Eq.(\ref{eck1}) as a function of coordinate $x$ and 
adiabatic parameter $A$. Vertical axes are scaled by $\frac{\hbar^2}{2m}$; Middle panel: Eckart potential for several adiabatic parameters. 
$A=1$ (black solid), $A=5$ (broken blue) and $A=10$ (dotted red). Lowest panel: Transmission probability in Eq.(\ref{transPR}) for the stationary tunneling as a function of $A$ in case of $k=1.2$. Length scale $l$=0.1 is used throughout in Figs.1-4. Units of space, time and other quantities used in Figs.1-8 are explained in the beginning of Section \ref{sample-tunn} and also below Eq.(\ref{eq3.24}). }
\end{figure}

\section{Examples}\label{sample-tunn}
We shall now investigate specific examples, and explicitly calculate the time-dependent transport coefficients 
in Eqs. (\ref{FFtransPR}) and (\ref{FFreflPR}) together with the driving electric field in Eq. (\ref{eq3.24}).
As typical examples of the stationary tunneling, we choose systems with (1) Eckart's potential \cite{eck} with tunable asymmetry and (2) double $\delta$-function barriers with tunable relative height \cite{read}. These systems are exactly solvable and allow one to evaluate both adiabatic and nonadiabatic contributions to transport coefficients during the fast forward dynamics. 

In our numerical analysis below, we shall use typical space and time units like
$L=10^{-2} \times$ {\it the linear dimension of a device} and $\tau=10^{-2} \times$
{\it the phase coherent time} and put $\frac{\hbar}{m}=1 (\times L^2\tau^{-1})$.
The above choice means that space coordinate $x$ (and other length parameters), time $t$, wavenumber $k$ and velocity  $v$ are scaled by
$L$, $\tau$, $L^{-1}$  and $L\tau^{-1}$, respectively.

\subsection{A system with Eckart's potential under adiabatically-tunable asymmetry}\label{Eckard}

\begin{figure}[htb]\label{Fig2}
\centering
\includegraphics[width=1.0\linewidth]{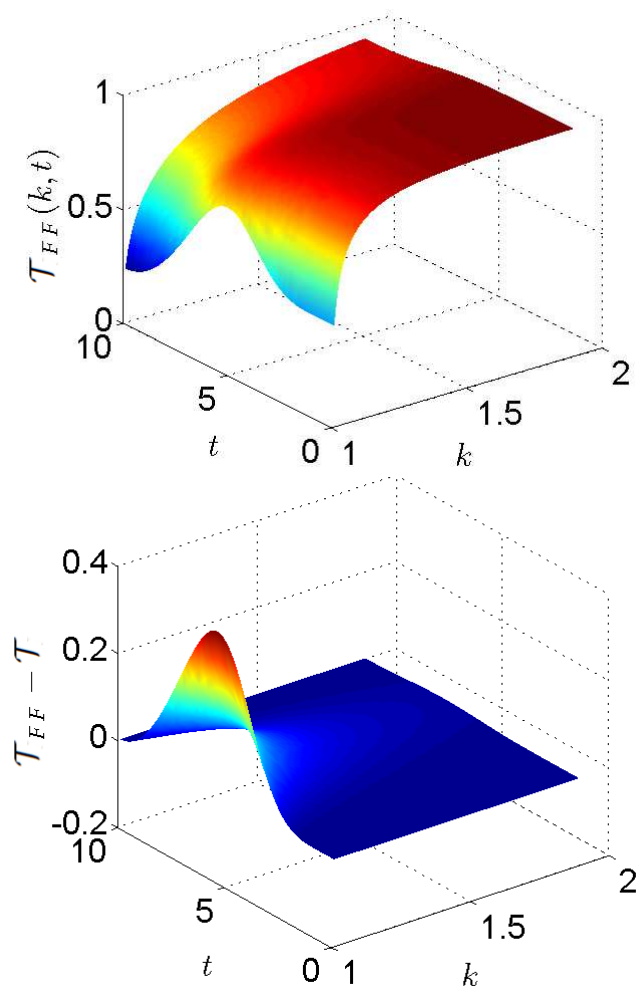}
\caption{\small $\mathcal{T}_{FF}(k,t)$ (upper panel) and its deviation from $\mathcal{T}(k,A(\Lambda(t)))$ (lower panel), 
as a function of wavenumber $k$ and time $t$. 
We choose $\bar{v}=1$ and $T_{FF}=10$ in the accelerated adiabatic parameter $A(\Lambda(t))$ in Eq.(\ref{eck11}), 
which are also used in Figs. 3 and 4.} 
\end{figure}

\begin{figure}[htb]\label{Fig3}
\centering
\includegraphics[width=1.0\linewidth]{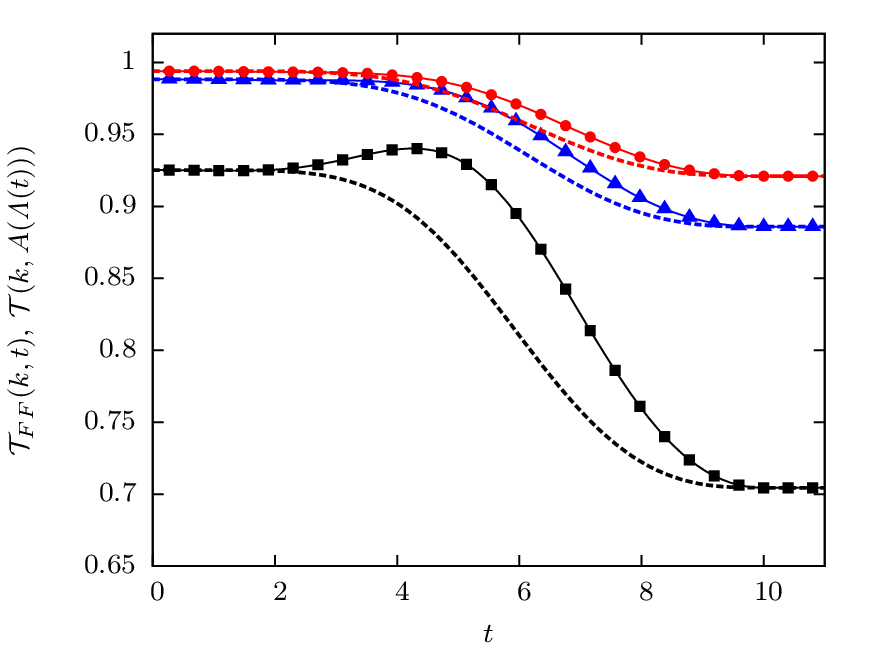}
\caption{Cross section of the upper panel of Fig.2 in the strip 
between $0.65 \le \mathcal{T}_{FF}(k,t) \le 1$ for input wavenumbers $k=1.2$ (black with squares), $1.6$ (blue with triangles) 
and $1.8$ (red with circles). Solid and broken lines correspond to $\mathcal{T}_{FF}(k,t)$ and $\mathcal{T}(k,A(\Lambda(t)))$, respectively.}
\end{figure}

\begin{figure}[htb]\label{elecEck}
\centering
\includegraphics[width=1.0\linewidth]{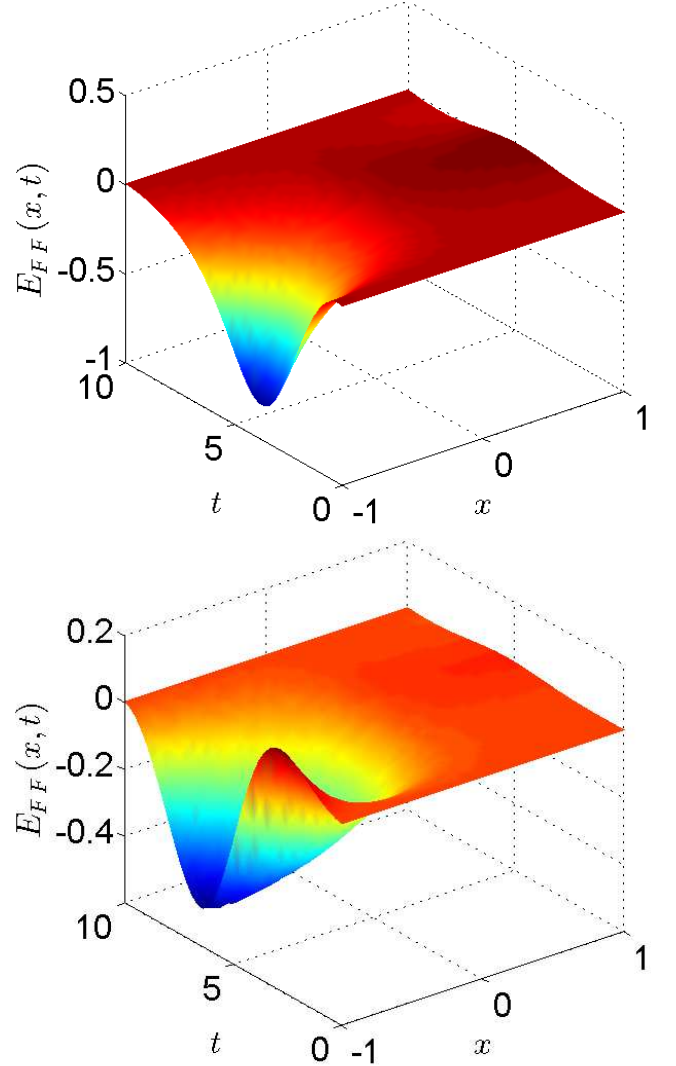}
\caption{\small Electric field as a function of space $x$ and time $t$ for wavenumbers $k=1.2$ (upper panel) and $1.8$ (lower panel).}
\end{figure}

This potential has a long history since the work by Eckart \cite{eck}, and has been used to describe the electron transmission through metal surfaces, nuclear reaction through a Coulomb barrier, etc. With use of length scale $l$, the potential is written as \cite{eck,vard}
\begin{eqnarray}\label{eck1}
V_{0}(x,A)=\frac{\hbar^2}{2m}\left(\frac{e^{x/l}}{1+e^{x/l}}+\frac{Ae^{x/l}}{(1+e^{x/l})^{2}}\right),
\end{eqnarray}  
which tends to $0$ and $\frac{\hbar^{2}}{2m}$ as $x\rightarrow-\infty$ and $x\rightarrow+\infty$, respectively. The 1st and 2nd terms on the right-hand side of Eq.(\ref{eck1}) are asymmetric and symmetric w.r.t. $x=0$, respectively. 
$A$ is the adiabatic parameter changing very slowly as
\begin{eqnarray}\label{eck2}
A=A(t)=\epsilon t,
\end{eqnarray}   
with $0<\epsilon\ll 1$. Figure 1 shows a profile of $V_{0}(x,A)$ as function of $x (|x|\leq10 l)$ and $A(0\leq A \leq10)$. $V_{0}$ has a maximum $V_{0}(x_{M}, A)=\frac{\hbar^{2}}{2m}\frac{(1+A)^{2}}{4A}$ at $x_{M}
=l \times \ln\left(\frac{A+1}{A-1}\right)$.

By making a variable change from $x$ to $\xi (=-\exp(x/l))$, the time-dependent Schr\"{o}dinger equation with Eckart's potential in Eq.(\ref{eck1}) becomes a differential equation for the Gauss' hypergeometric function $F$. 
Then the exact solution for electronic wave function is given by \cite{eck,vard} 
\begin{eqnarray}\label{eck3}
\phi_{0}(x,k,A)=t_r(1-\xi)^{\textit{i}k'l}\left(-\frac{\xi}{1-\xi}\right)^{ikl}\quad\quad\quad\nonumber\\
\times F\left(\frac{1}{2}+\textit{i}(k-k'+\delta)l,\frac{1}{2}+\textit{i}(k-k'-\delta)l,\right.\nonumber\\1-2ik'l,
\left.\frac{1}{1-\xi}\right),\quad\quad\quad\quad\quad\quad\quad\quad\quad\quad\quad
\end{eqnarray}
with 
\begin{align}\label{eck4}
\begin{split}
k^{2}&=\frac{2mE}{\hbar^{2}}, \\
k'^{2}&=k^{2}-1, \\
\delta&=\sqrt{A-\frac{1}{4l^2}},\\
t_r &= \frac{\Gamma(\frac{1}{2}+i(-k-k'-\delta)l)\Gamma(\frac{1}{2}+i(-k-k'+\delta)l)}{\Gamma(1-2ik'l)\Gamma(-2ikl)}. 
\end{split}
\end{align}
We should note that the adiabatic parameter $A$ shows up through $\delta$ in Eq.(\ref{eck4}). In Eqs. (\ref{eck3}) 
and (\ref{eck4}), we have corrected the mistakes included in \cite{eck}, which was pointed out in \cite{vard}.

We can use  the linear transformation formula among Gauss'  hypergeometric functions \cite{abram},
which is convenient to see the asymptotic behavior in the region $x\rightarrow-\infty(\xi\rightarrow0)$. In fact, we see there a sum of
 the incoming and reflective waves as 
\begin{eqnarray}\label{eck6}\phi_{0}=e^{ikx}+r_f e^{-ikx}.
\end{eqnarray}
In the opposite asymptotic region $x\rightarrow\infty(\xi\rightarrow-\infty)$, $\phi_{0}$ in Eq.(\ref{eck3}) becomes a transmitting wave:
\begin{eqnarray}\label{eck5}
\phi_{0}=t_r(-\xi)^{ik'l}=t_r\exp(ik'x).
\end{eqnarray}


In case of $Al^2<\frac{1}{4}$, the transition probabilitiy becomes: 
\begin{eqnarray}\label{eck8}
\mathcal{T}(k,A)&=&\frac{k'}{k}\left|t_r\right|^{2} \nonumber\\
&=&\frac{\cosh(2\pi(k+k')l)-\cosh(2\pi(k-k')l)}{\cosh(2\pi(k+k')l)+\cos(2\pi|\delta| l)}. \nonumber\\
\end{eqnarray} 
In case of $Al^2\geq\frac{1}{4}$, on the other hand, we have 
\begin{eqnarray}\label{eck9}
\mathcal{T}(k,A)=\frac{\cosh(2\pi(k+k')l)-\cosh(2\pi(k-k')l)}{\cosh(2\pi(k+k')l)+\cosh(2\pi\delta l)}. \nonumber\\
\end{eqnarray}
The reflection probability is given by 
\begin{eqnarray}\label{eck10}\mathcal{R}(k,A)=1-\mathcal{T}(k,A).\end{eqnarray}

In the fast forward of the adiabatic dynamics, the standard time $t$ is replaced by the advanced time $\Lambda(t)=\int^{t}_{0}\alpha(t')dt'$, and taking the limit $\bar{\alpha}\rightarrow\infty$, $\epsilon\rightarrow0$ with $\bar{\alpha}\varepsilon=\bar{\textit{v}}$ kept constant, the accelerated adiabatic parameter is now given by
\begin{eqnarray}\label{eck11}
A(\Lambda(t))=\bar{\textit{v}}
\left(t-\frac{T_{FF}}{2\pi}\sin\left(\frac{2\pi}{T_{FF}}t\right)\right),
\end{eqnarray}
as given in Eq.(\ref{accel-ad}). Then, using Eqs. (\ref{FFtransPR}) and (\ref{FFreflPR}), $\mathcal{T}_{FF}(k,t)$ and
$\mathcal{R}_{FF}(k,t)$ can be computed.

If we shall confine to the parameter region $0\le A \le 10$ and employ the length scale $l=0.1$ as in Fig.1, we see the saturation of the potential $V_0(x,A)$ for $x\le -1$ and $x\ge 1$, as
\begin{align}\label{eck-app}
\begin{split}
&|V_0(x,A)| \le 10^{-3}    \qquad \qquad \quad{\rm for} \qquad x\le-1, \\
&|V_0(x,A)-\frac{\hbar^2}{2m}|\le 10^{-3}  \qquad {\rm for} \qquad  x\ge 1.
\end{split}
\end{align}
Then the stationary values $\mathcal{T}(k,A)$ and $\mathcal{R}(k,A)$  do not depend on the choice of $x_2$ and $x_1$ so long as
$x_2\ge 1$ and $x_1\le -1$. Therefore, in our numerical calculation of  $\mathcal{T}_{FF}(k,t)$ and
$\mathcal{R}_{FF}(k,t)$ in Eqs. (\ref{FFtransPR}) and (\ref{FFreflPR}), we take $x_1=-1$ and $x_2=1$.
As for the lower limit of the integration there, we choose $c=0$.

Figure 2 shows both $\mathcal{T}_{FF}(k,t)$ and its deviation from the stationary counterpart $\mathcal{T}(k,A(\Lambda(t)))$ as a function of $k(1\leq k \leq 2)$ 
and $t(0\leq t \leq T_{FF})$. $\mathcal{T}_{FF}$  shows deviation from $\mathcal{T}(k,A(\Lambda(t)))$, but agrees with the latter at $t=T_{FF}$ for any input wavenumbers $k$. Figure 3 is a cross section of the upper panel of Fig.2 for  several input wavenumbers $k$, showing that $\mathcal{T}_{FF}(k,t)$ 
recovers the stationary value at $t=T_{FF}$.

The electric field $E_{FF}$ to guarantee the fast forward is calculated with use of the formula in Eq.(\ref{eq3.24}), where $\partial_x\eta$ is available from Eq.(\ref{eck3}) and   $\partial_x\theta$ is calculable from Eq.(\ref{dtheta}) together with Eq.(\ref{eck3}). Figure 4 shows the $3D$ plots of $E_{FF}$ as a function 
of $x(|x|<1)$ and $t(0\leq t\leq T_{FF})$ for several input wavenumbers $k$.
In SI unit for electric field, typical absolute value $E_{FF}=0.5$ in ordinates in Fig. 4 means $E^{FF}_{SI}=5 \times 10^{11}$
in case of IR lasers of wave length $\sim$ 1$\mu$m (see the argument  below Eq. (\ref{eq3.24})).

\subsection{Double $\delta$-function barriers with adiabatically-tunable asymmetry}

\begin{figure}[htb]\label{pt-DD}
\centering\includegraphics[width=1.0\linewidth]{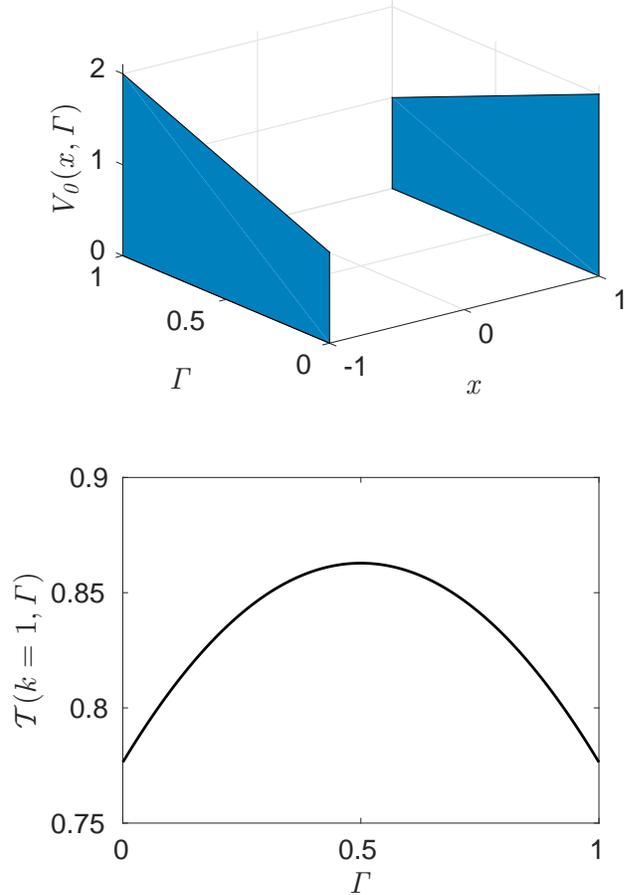}
\caption{Upper panel: Asymmetric potential consisting of double $\delta$-functions in Eq.(\ref{eq.b.v}), 
as a function of space variable $x$ and adiabatic parameter $\Gamma$. Vertical axis is scaled by $\frac{\hbar^2}{2m}$;   Lower panel: Transmission probability in Eq.(\ref{transPR})  for the stationary tunneling in case of $k=1$. $a=1$ is used throughout in Figs.5-8.}
\end{figure}

We shall move to analyze another example: the fast-forward of adiabatic control of double $\delta$-function barriers with tunable asymmetry, which is a simplified variant of the double barrier in semiconductor heterostructures. Assuming the barriers located at $x=\pm a$, the underlying Hamiltonian is given by
\begin{eqnarray}\label{eq.ham.b}
\hat{H}_{0}=-\frac{\hbar^{2}}{2m}\frac{d^{2}}{dx^{2}}+V_{0}(x,\Gamma).
\end{eqnarray}

Here
\begin{eqnarray}\label{eq.b.v}
V_{0}(x,\Gamma)=\frac{\hbar^2}{2m}\left[(h_{min}+\Gamma) \delta(x+a)+(h_{max}-\Gamma)\delta(x-a)\right], \nonumber\\
\end{eqnarray}
with $\Gamma$ the adiabatic parameter defined by
\begin{eqnarray}\label{eq.b.g}
\Gamma=\Gamma(t)=\varepsilon t\quad\quad (\varepsilon\ll 1),
\end{eqnarray}
which is assumed to vary from $\Gamma(0)=0$ to
$\Gamma(T)=h_{max}-h_{min}\equiv \Delta h$ with $T=\frac{\Delta h}{\varepsilon}\gg1$.

Figure 5 shows a profile of the potential as a function of $x(-a-0\leq x\leq a+0)$ with $a=1$ and $\Gamma$($0\leq\Gamma\leq\Delta h)$ with $h_{max}=2, h_{min}=1$ and $\Delta h=1$. 

Firstly, we consider the stationary tunneling state available from the time-independent Schr$\ddot{\textrm{o}}$dinger equation
\begin{eqnarray}\label{eq.b.sch}
\hat{H}_{0}(\Gamma)\phi_{0}=E(\Gamma)\phi_{0}.
\end{eqnarray}
Let's define 3 domains, $D_{L}(:  x < -a), D_{C}(: -a \le x \le a)$ and $D_{R}(: x>a)$ and suppose the wave-functions, respectively, as
\begin{align}\label{eq.b.wf}
\begin{split}
\phi^{L}_{0}&= e^{ikx}+r_f(\Gamma)e^{-ikx} \qquad\quad (\textrm{in} \quad D_{L}), \\
\phi^{C}_{0}&= A(\Gamma)e^{ikx}+B(\Gamma)e^{-ikx}\quad (\textrm{in} \quad D_{C}), \\
\phi^{R}_{0}&= t_r(\Gamma)e^{ikx}   \quad\quad\quad\quad\quad\quad\quad(\textrm{in} \quad D_{R}),
\end{split}
\end{align}
where $\phi_0^{L}$ is a sum of the incident and reflective wave-functions. $r_f (t_r)$ means reflection (transmission) coefficient which is complex.

Unknown coefficients $A, B, r_f$ and $t_r$ can be obtained by using two constraints: (1) the continuity of the wave-function 
$\phi_0$ at $x=\pm a$; (2) the finite discontinuity of the derivative, $\frac{d}{dx}\phi$, available from the  local integration of Eq.(\ref{eq.ham.b}) in the vicinity of $x=\pm a$. With prescription of $\frac{\hbar^2}{2m}=1$, the results are \cite{read}:
\begin{align}\label{eq.b.t}
\begin{split}
\Delta(k)&=(h_{min}+\Gamma)(h_{max}-\Gamma)(-1+e^{4iak})\qquad \qquad \qquad \qquad\ \\
&+4k^{2}+2i(h_{min}+h_{max})k,\qquad
\\
t_r(\Gamma)&=\frac{4k^{2}}{\Delta(k)}, \qquad \qquad  \qquad \qquad \qquad \qquad \qquad \qquad\qquad \qquad
\\
r_f(\Gamma)&=\frac{e^{2iak}}{\Delta(k)}\left\{(h_{min}+\Gamma)(h_{max}-\Gamma)(-1+e^{-4iak})\right.\qquad \qquad
\\
&\left.-2ik(h_{max}-\Gamma+(h_{min}+\Gamma) e^{-4ika})\right\},\qquad \\
A(\Gamma)&=\frac{2k(2k+i(h_{max}-\Gamma))}{\Delta(k)},\qquad \qquad \qquad \qquad \qquad \qquad \\
B(\Gamma)&=\frac{-2ik(h_{max}-\Gamma)e^{2iak}}{\Delta(k)}.\qquad \qquad \qquad \qquad\qquad \qquad
\end{split}
\end{align}
\begin{figure}[htb]\label{trnsDD}
\centering
\includegraphics[width=1.0\linewidth]{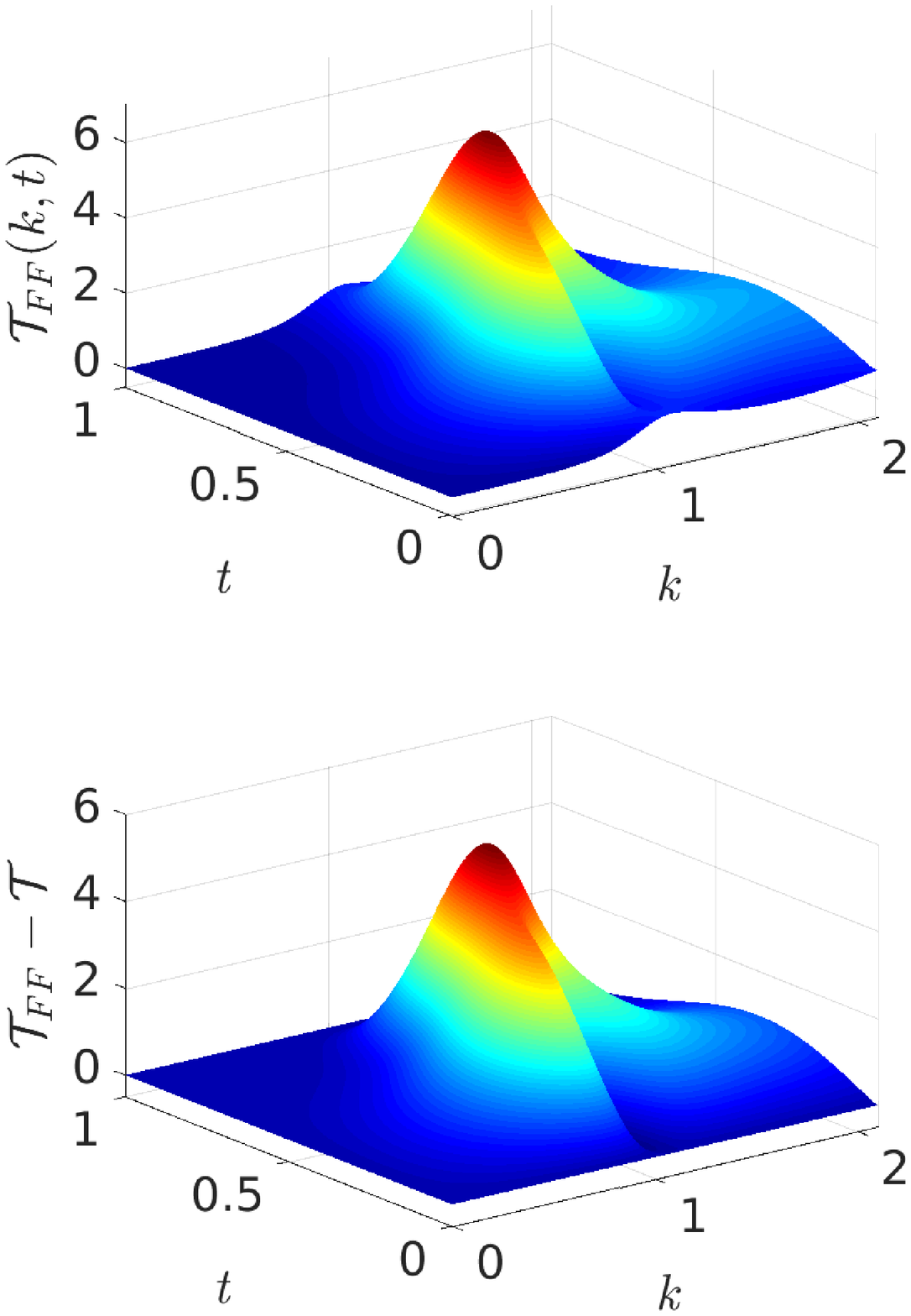}
\caption{\small $\mathcal{T}_{FF}(k,t)$ (upper panel) and its deviation from $\mathcal{T}(k,\Gamma(\Lambda(t)))$ (lower panel), as a function of wavenumber $k(0\leq k\leq2)$ and time $t$. We choose $\bar{v}=1$ and $T_{FF}=1$ in the accelerated adiabatic parameter $\Gamma(\Lambda(t))$, 
which are also used in Figs. 7 and 8.} 
\end{figure} 

In the fast-forward of the adiabatic dynamics, the time $t$ in $\Gamma(t)$ is replaced by $\Lambda=\int^{t}_{0}\alpha(t')dt'$ and we take the limit $\bar{\alpha}\rightarrow\infty$, $\varepsilon\rightarrow0$ with $\bar{\alpha}\varepsilon=\bar{v}$ fixed. Then the accelerated adiabatic parameter 
$\Gamma(\Lambda(t))$ has the same form as $A(\Lambda(t))$ in Eq.(\ref{eck11}). 

Having recourse to the formulas in Eqs.(\ref{FFtransPR}) and (\ref{FFreflPR}), we can calculate $\mathcal{T}_{FF}(t)$ and $\mathcal{R}_{FF}(t)$ at $x_2=a+0$ before the right barrier and at $x_1=-a-0$ behind the left barrier, respectively. To evaluate the nonadiabatic correction in Eqs.(\ref{FFtransPR}) and (\ref{FFreflPR}), we again choose $c=0$ as the lower limit of integrations and use the following result of integrations:
\begin{eqnarray}\label{eq.b.g1}
&& J^{(\pm a)} \equiv \int_0^{(\pm a)} \partial_{\Gamma}{\bar{\phi}}^{2}_{0} dx \nonumber\\
&&=\partial_{\Gamma} \left(\pm a(\bar{A}^2+\bar{B}^2)+\frac{2\bar{A}\bar{B}}{k}\sin(\pm ak)\cos(\pm a k+\alpha-\beta)\right),
\nonumber\\
\end{eqnarray}
where $\bar{A}(\bar{B})$ is the real positive amplitude and $\alpha(\beta)$ is the phase of the complex coefficients 
$A(\Gamma) (B(\Gamma))$ defined in Eq.(\ref{eq.b.t}). In Eq.(\ref{eq.b.g1}), $+$ and $-$ in the sign $\pm$ correspond to $x_2$ and $x_1$, respectively.

\begin{figure}[!h]\label{Fig7}
\centering
\includegraphics[width=1.0\linewidth]{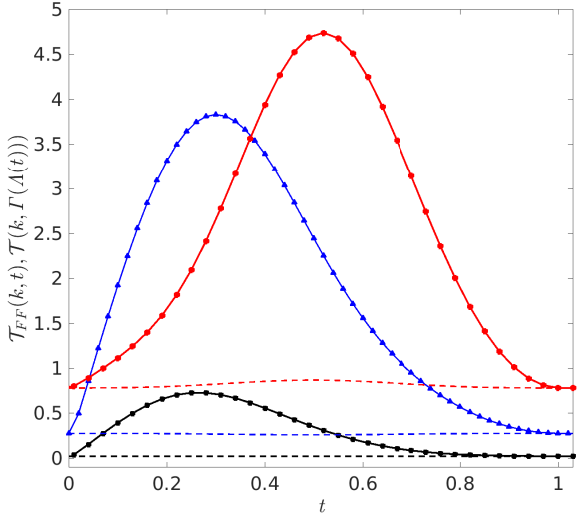}
\caption{\small Cross section of the upper panel of Fig.6 for input wavenumbers $k=0.4$ (black with squares), $0.8$ (blue with triangles) and 
$1.2$ (red with circles).  Solid and broken lines correspond to
$\mathcal{T}_{FF}(k,t)$ and $\mathcal{T}(k,\Gamma(\Lambda(t)))$, respectively.} 
\end{figure}

\begin{figure}[!h]\label{elecDD}
\centering
\includegraphics[width=1.0\linewidth]{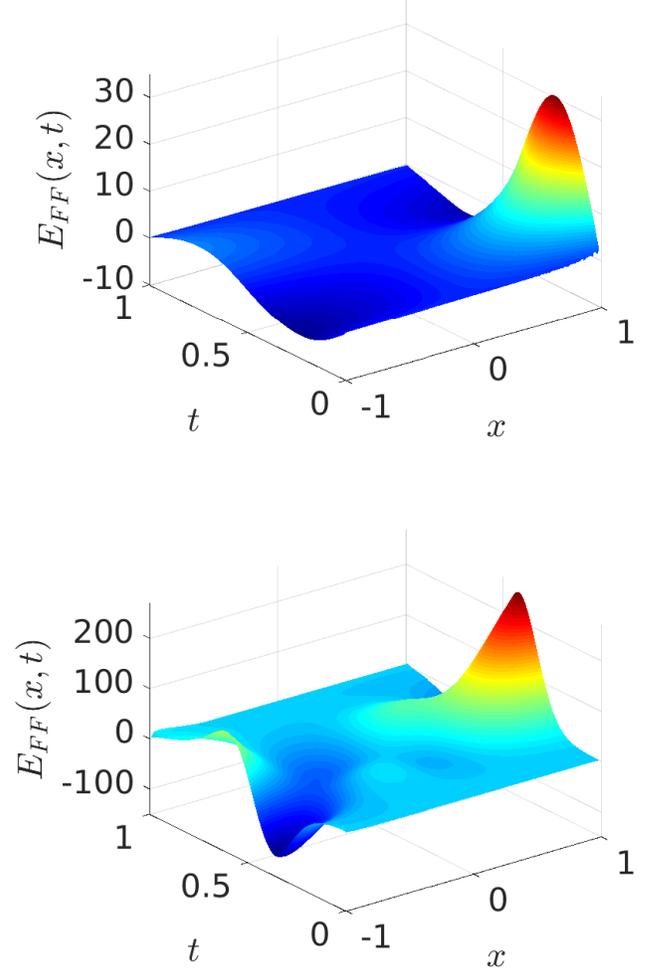}
\caption{\small Electric field as a function of space $x$ and time $t$ for wavenumbers $k=0.4$ (upper panel) and $1.2$ (lower panel).} 
\end{figure}

Figure 6 shows both $\mathcal{T}_{FF}(k, t)$ (upper panel) and its deviation from the stationary counterpart $\mathcal{T}(k, \Gamma(\Lambda(t)))$ (lower panel) 
as a function of $k(0\leq k\leq2)$ and $t(0\leq t\leq T_{FF})$. 
$\mathcal{T}_{FF}$ shows to reach the stationary value at $t=T_{FF}$.  Figure 7 is a cross section of the upper panel of Fig.6  for several input wavenumbers $k$, showing that $\mathcal{T}_{FF}(k,t)$ recovers the stationary value at $t=T_{FF}$. 
The large deviation of  $\mathcal{T}_{FF}(k, t)$ from its stationary counterpart in Figs. 6 and 7 is caused by the driving electric field  which is stronger in the case of  double $\delta$-function barriers than in the case of Eckart's potential (see Fig. 8).

The electric field $E_{FF}$ which guarantees the fast forward can be evaluated with use of Eq.(\ref{eq3.24}). Here $\partial_{x}\eta$ is available from the wavefunction in each domain in Eq.(\ref{eq.b.wf}). On the other hand, 
$\partial_{x}\theta$ in Eq.(\ref{dtheta}) can be available from the following results of the integration
\begin{eqnarray}\label{eq.4.31}
&& J^{(x)}  \equiv  \int_0^{x} \partial_{\Gamma}{\bar{\phi}}^{2}_{0} dx^{\prime}
 \nonumber\\
&=&
\begin{cases}
J^{(-a)}+\partial_{\Gamma} \left((1+\bar{r}^2)(x+a)\right.\\
\left. +\frac{2\bar{r}}{k}\sin(k(x+a)\cos(k(x-a)-\gamma)\right)& (x \quad \textrm{in} \quad D_{L}),
 \\
 \\
\partial_{\Gamma} \left((\bar{A}^2+\bar{B}^2)x \right.\\
\left.+\frac{2\bar{A}\bar{B}}{k}\sin(kx)\cos(kx+\alpha-\beta)\right)& (x \quad \textrm{in} \quad D_{C}),
\\
\\
J^{(a)}+\partial_{\Gamma} \left(\bar{t}^2(x-a)\right.\\
\left.+\frac{2\bar{t}}{k}\sin(k(x-a)\cos(k(x+a)+\tau)\right)& (x \quad \textrm{in} \quad D_{R}),
\end{cases}\nonumber\\
\end{eqnarray}
where $\bar{r}(\bar{t})$ is the real positive amplitude and $\gamma(\tau)$ is the phase of the complex coefficients 
$r_f(\Gamma) (t_r(\Gamma))$ defined in Eq.(\ref{eq.b.t}). 
Figure 8 shows $E_{FF}$ as a function of $t$ and $x$ in the range $0 \le t\le T_{FF}$ and $|x| \le 1$ for several input wavenumbers $k$.
In SI unit for electric field, typical absolute value $E_{FF}=100$ in ordinates in Fig. 8 means $E^{FF}_{SI}=10^{14}$
in case of IR lasers of wave length $\sim$ 1$\mu$m. 
The localized high peaks and deep dips arise when $\bar{\phi}_0^2$ in the denominator on the right-hand side of Eq.(\ref{dtheta}) takes small but non-zero values due to the interference between a pair of waves in the domain $D_C$ in Eq.(\ref{eq.b.wf}) that forms an internal structure, i.e., a potential well surrounded by a pair of barriers.

Numerical results in this Section convey some basic features of  the fast-forward observation of the transport coefficients under the adiabatically-changing barrier. The results will be more-or-less modified by varying the mean time-scaling factor $\bar{v}$,
the spatial size of barriers relative to wave length of the incoming particle, etc., which should be investigated separately in due course.

\section{Conclusion}\label{concl}
We have proposed a scheme of the exact fast forward of adiabatic control of stationary tunneling states with use of  the electromagnetic field, which allows the fast forward with complete fidelity, namely the exact acceleration of both the amplitude and phase of wave functions throughout the fast-forward time range.
For the incoming particle with fixed energy, the scheme realizes the fast-forward observation of transport coefficients under the adiabatically-changing barrier. The fast-forwarded transport coefficients are decomposed into the adiabatic part which satisfies the unitarity and the nonadiabatic one which vanishes only at the end of the fast forwarding. We have also elucidated the modulation of the phase of complex scattering coefficients.

As typical examples we have investigated systems with (1) Eckart's potential with tunable asymmetry and (2) double $\delta$-function barriers under tunable relative height. The driving  electric field is evaluated to guarantee the stationary tunneling state during a rapid change of the barrier.
The nonadiabatic contribution to transport coefficients proves to be remarkable in case that barriers have internal structures. Detailed numerical analysis of the dependence on the mean time-scaling factor $\bar{v}$, the spatial size of barriers relative to wave length of the incoming particle, etc. will constitute a future independent subject.  The present scheme will be a promising extension of the fast forward of adiabatic dynamics of the bound ground states to that of open tunneling states.

\begin{acknowledgments}
We are grateful to Zarif Sobirov, Yuki Izumida and Makoto Hosoda for their critical comments.
\end{acknowledgments}

\appendix
\section{Gauge transformation between systems with complete and incomplete fidelities}\label{app1}
In the context of fast forward of adiabatic dynamics of bound states, the scheme presented here is compatible with the one in Refs.\cite{mas2,mas3}.
Let us introduce the gauge transformation into Eqs.(\ref{ffeq}), (\ref{eq3.17}), and (\ref{eq3.19}) (with the dynamical factor replaced by
$e^{-\frac{i}{\hbar}\int_0^t E(R(\Lambda(s)))ds}$)
 as follows
\begin{align}\label{eq3.20}
\begin{split}
\psi_{FF}&\rightarrow \psi^{MN}_{FF}e^{\textit{if}},\\
V_{FF}&\rightarrow  V^{MN}_{FF}-\frac{\hbar}{q}\partial_{t}\textit{f},\\
\textbf{A}_{FF}&\rightarrow \textbf{A}^{MN}_{FF}+\frac{\hbar}{q}\nabla\textit{f}
\end{split}
\end{align}
with the phase defined by
\begin{eqnarray}\label{eq3.21}
f=-v(t)\theta(\textbf{x},R(\Lambda(t))).
\end{eqnarray}
Then we find
\begin{align}\label{eq3.22}
\begin{split}
\psi^{MN}_{FF}&=\bar{\phi}_{0}(\textbf{x},R(\Lambda(t)))
e^{\textit{i}\eta(\textbf{x},R(\Lambda(t)))}e^{\textit{i}v(t)\theta(\textbf{x},R(\Lambda(t)))} \\
&\times e^{-\frac{i}{\hbar}\int_0^t E(R(\Lambda(s)))ds},\\
V_{FF}^{MN}&=-\frac{\hbar^{2}}{m}v(t)\nabla\theta\cdot\nabla\eta-\frac{\hbar^{2}}{2m}(v(t))^{2}(\nabla\theta)^{2} \\
&-\hbar v(t)\partial_{R}\eta-\hbar\dot{v}(t)\theta-\hbar(v(t))^{2}\partial_{R}\theta, \\
\textbf{A}_{FF}^{MN}&=0,
\end{split}
\end{align}
and $\psi^{MN}_{FF}$ proves to satisfy
\begin{eqnarray}\label{eq3.23}
\textit{i}\hbar\frac{\psi^{MN}_{FF}}{\partial t}=\left(-\frac{\hbar^{2}}{2m}\nabla^2+V_{0}+qV^{MN}_{FF}\right)\psi^{MN}_{FF}.
\end{eqnarray}
 Eqs.(\ref{eq3.22}) and (\ref{eq3.23}) together with notion of $q=1$ reproduces the preceding issue~\cite{mas2,mas3} which generated the exact adiabatic state only at the final time $t=T_{FF}$, but failed to keep the perfect fidelity in the intermediate time range $0<t<T_{FF}$. 

In fast forward of the particular adiabatic control of bound states, $V_{FF}^{MN}$ in Eq.(\ref{eq3.22}) has an expression convenient to generate the counter-diabatic potential \cite{dr1,dr2,mb}, which we shall briefly explain below.

Consider the original potential controlled by the scale-invariant adiabatic 
expansion and contraction \cite{kle,campo,djc}, as given by
\begin{align}
\label{eqa5}
V_0=\frac{1}{R^2}U_0\left(\frac{x}{R}\right),
\end{align}
where $R$ is the adiabatic parameter as in Eq.(\ref{eq3.1}). The corresponding $1D$ eigenvalue problem for bound systems yields ground and excited states whose normalized forms are commonly given by
\begin{align}
\label{eqa6}
\phi_0=\frac{1}{\sqrt{R}}f \left(\frac{x}{R}\right),
\end{align}
where $f=\bar{f}e^{i\eta}$ with real amplitude $\bar{f}$ and phase $\eta$.
Then, with use of a new variable $X\equiv \frac{x}{R}$, Eq.(\ref{dtheta}) becomes
\begin{align}
\label{eqa7}
\partial_x \theta=-\frac{m}{\hbar} \frac{R}{|\bar{f}(X)|^2} 
\partial_R \int^X |\bar{f}(X^{\prime})|^2 d X^{\prime}.
\end{align}
Here the indefinite integral is used because the lower limit of integration is arbitrary. Noting $\partial_R=\frac{\partial X }{\partial R} \frac{\partial }{\partial X}=-\frac{x}{R^2} \frac{\partial }{\partial X}$, Eq.(\ref{eqa7}) reduces to
\begin{align}
\label{eqa8}
\partial_x \theta= \frac{m}{\hbar} \frac{x}{R} \frac{|\bar{f}(X)|^2}{|\bar{f}(X)|^2}=\frac{m}{\hbar R} x.
\end{align}
In the second equality of Eq.(\ref{eqa8}), we prescribed $\lim_{X\rightarrow X_c}\frac{|\bar{f}(X)|^2}{|\bar{f}(X)|^2}=1$ if $\bar{f}(X)$ will be $\bar{f}(X_c)=0$ at $X=X_c$. From Eq.(\ref{eqa8}), one finds \cite{mas3}:
\begin{align}
\label{eqa9}
\begin{split}
\theta&=\frac{m}{2\hbar R} x^2, \\
\partial_R \theta&=-\frac{m}{2\hbar R^2} x^2.
\end{split}
\end{align}
In the simple case that $\phi_0$ in Eq.~(\ref{eqa6}) is real, i.e., $\eta=0$,  $V_{FF}^{MN}$ in Eq.(\ref{eq3.22}) becomes
\begin{align}
\label{eqa10}
V_{FF}^{MN}=-\frac{m\ddot{R}}{2R}x^2,
\end{align}
where $R=R(\Lambda(t))$, $v(t)=\dot{R}$ and $\dot{v}(t)=\ddot{R}$ in Eq.(\ref{accel-ad}) are used. $V_{FF}^{MN}$ in Eq.(\ref{eqa10}) is nothing but the counter-diabatic potential in the scale-invariant bound systems~\cite{campo,djc}. The generalization of the above argument to the case which includes the scale-invariant adiabatic translation is straightforward. 

Thus the fast forward approach \cite{mas1,mas2,mas3} applied to the scale-invariant bound systems is free from the problem of nodes, although such a problem might appear when we shall manage excited states of the bound systems that break the scale invariance. On the other hand, as explained around Eq. (\ref{stcurr}), the stationary (or steady) tunneling state investigated in the present paper has no nodes and is free from both the problem of nodes and the constraint of scale invariance.

\section{Derivation of the driving $A_{FF}$ and $V_{FF}$ potentials in Eqs.(\ref{eq3.15}) and (\ref{eq3.16})}\label{app2}
As for space derivatives of $\tilde{\phi}_0^{reg}$ in Eq.(\ref{eq3.13}), 
we shall have recourse to the formulas: $\textrm{Re}\left[\frac{\partial_x\tilde{\phi}^{reg}_{0}}{\tilde{\phi}^{reg}_{0}}\right]
=\partial_x(\ln\bar{\phi}_{0})$, $\textrm{Im}\left[\frac{\partial_x\tilde{\phi}^{reg}_{0}}{\tilde{\phi}^{reg}_{0}}\right]
=\partial_x\eta
+\epsilon\partial_x\theta$, $\textrm{Re}\left[\frac{\partial_x^{2}\tilde{\phi}^{reg}_{0}}{\tilde{\phi}^{reg}_{0}}\right]=\frac{\partial_x^{2}\bar{\phi}^{reg}_{0}}{\bar{\phi}_{0}}-(\partial_x\eta+\epsilon\partial_x\theta)^{2}=\frac{2m}{\hbar^{2}}(V_{0}-E)-2\epsilon\partial_x\eta\cdot\partial_x\theta-\epsilon^{2}(\partial_x\theta)^{2},$ $\textrm{Im}\left[\frac{\partial_x^{2}\tilde{\phi}^{reg}_{0}}{\tilde{\phi}^{reg}_{0}}\right]=\frac{2\partial_x\bar{\phi}_{0}}{\bar{\phi}_{0}}(\partial_x\eta+\epsilon\partial_x\theta)+(\partial_x^{2}\eta+\epsilon\partial_x^{2}\theta)=\epsilon \left(\partial_x^{2}\theta+2\partial_x(\ln\bar{\phi}_{0})\cdot\partial_x\theta \right).$
In obtaining the final issue in each of the last two equations, we used the identities, 
\begin{align}\label{eta-equality}
\begin{split}
\frac{\partial_x^{2}\bar{\phi}_{0}}{\bar{\phi}_{0}}&-(\partial_x\eta)^{2}=\frac{2m}{\hbar^{2}}(V_{0}-E),\\
\partial_x^{2}\eta&+2\frac{\partial_x\bar{\phi}_{0}}{\bar{\phi}_{0}}\cdot\partial_x\eta=0,
\end{split}
\end{align}
which are available from the adiabatic eigenvalue problem in Eq.(\ref{eq3.3}) for the stationary state in Eq.(\ref{eq3.8}). 

Equation (\ref{aff}) now becomes
\begin{eqnarray}\label{eq3.14}
\bar{\phi}^{2}_{0}\partial_xA_{FF}&+&2\bar{\phi}_{0}\partial_x\bar{\phi}_{0}\cdot A_{FF} \nonumber\\
&+&\hbar(\alpha-1)\epsilon(\bar{\phi}^{2}_{0}\partial_x^{2}\theta+2\bar{\phi}_{0}\partial_x\bar{\phi}_{0}\cdot \partial_x\theta)=0,\nonumber\\
\end{eqnarray}
which is found to be satisfied by $A_{FF}$ in Eq.(\ref{eq3.15}).
Using Eq.(\ref{eq3.15}) together with spatial derivatives of $\tilde{\phi}_0^{reg}$ described above Eq.(\ref{eta-equality}),
$V_{FF}$ in Eq.(\ref{vff}) turns out to take the form in Eq.(\ref{eq3.16}).

\section{Analysis of continuity equation of the fast-forward dynamics}\label{anal-cont}
Equation (\ref{FF-nonunitari})
is also available from the continuity equation of the fast-forward dynamics:
\begin{equation}\label{FF-conserv}
\partial_t|\psi_{FF}|^2+\partial_x j_{FF}(x,t)=0,
\end{equation}
where $|\psi_{FF}|^2=\bar{\phi}_0^2(R(\Lambda(t)))$. By integrating Eq.(\ref{FF-conserv}) from $x=x_1$
to $x=x_2$ and using $\partial_t=\frac{dR}{dt}\partial_R=v(t)\partial_R$, we have
\begin{eqnarray}\label{check-nonunitari}
j_{FF}(x=x_2, t)&-&j_{FF}(x=x_1, t)\nonumber\\
&=&-v(t)\int_{x_1}^{x_2}\partial_R\bar{\phi}_0^2dx.
\end{eqnarray}
Dividing the equality in Eq.(\ref{check-nonunitari}) by $j_0(=\frac{\hbar}{m}k)$,
we can confirm Eq.(\ref{FF-nonunitari}).

\end{document}